\documentclass{PoS}
\usepackage{ulem}
\usepackage{algorithm}
\usepackage{algorithmic}

\rightline{\sffamily HUPD-1006}\vspace*{-10mm}

\title{Domain Decomposition method on GPU cluster}

\ShortTitle{Domain Decomposition method on GPU cluster}

\author{\speaker{Yusuke Osaki} and Ken-Ichi Ishikawa\\
        Graduate School of Science, Hiroshima University, 
        Kagamiyama 1-3-1, Higashi-Hiroshima, 739-8526, Japan\\
        E-mail: \email{ozaki@theo.phys.sci.hiroshima-u.ac.jp}\\
        E-mail: \email{ishikawa@theo.phys.sci.hiroshima-u.ac.jp}}

\abstract{
Pallalel GPGPU computing for lattice QCD simulations has a bottleneck on the GPU to GPU data communication due to the lack of the direct data exchanging facility.
In this work we investigate the performance of quark solver using the restricted additive Schwarz (RAS) preconditioner on a low cost GPU cluster.
We expect that the RAS preconditioner with appropriate domain-decomposition and task distribution reduces the communication bottleneck.
The GPU cluster we constructed is composed of four PC boxes, two GPU cards are attached to each box, and we have eight GPU cards in total.
The compute nodes are connected with rather slow but low cost Gigabit-Ethernet.
We include the RAS preconditioner in the single-precision part of the mixed-precision nested-BiCGStab algorithm and the single-precision task is distributed to the multiple GPUs.
The benchmarking is done with the $O(a)$-improved Wilson quark on a randomly generated gauge configuration with the size of $32^4$.
We observe a factor two improvment on the solver performance with the RAS precoditioner compared to that without the preconditioner and find that the improvment mainly comes from the reduction of the communication bottleneck as we expected.
}

\FullConference{The XXVIII International Symposium on Lattice Field Theory, Lattice2010\\
		June 14-19, 2010\\
		Villasimius, Italy}

\begin{document}

\section{Introduction} 
\label{sec:Intro}

The application of General-Purpose GPU (GPGPU) computing for lattice QCD simulations is very attractive and there have been several studies in the literature~\cite{Egri:2007,Clark:2010}.
The most of the previous GPGPU works for lattice QCD simulations have focused on the acceleration of the quark solver using a single GPU card.
However single GPU is not sufficient to simulate QCD with more realistic lattice parameters, such as over $32^4$ lattices with physical quark masses, due to the lack of memory size and required sustained speed.
Thus we need parallel GPGPU computing platforms with multiple GPU cards.
This year several studies for lattice QCD simulations with/without multiple GPU cards are reported in this conference~\cite{Lat2010GPU}.
In this paper we report our trial and benchmarking study of the quark solver on a GPU cluster we developed.

One of the bottleneck of parallel computing is in the data communication among compute processing units generally.
Any multipile GPU system such as PC cluster with GPUs (= GPU cluster) also has the same bottleneck.
The situation is worse for the GPU cluster since there is no on-board facility to directly exchange the data between GPU memories on distinct nodes.
Typically the data path consists of the PCI-Express path between host CPU memory and GPU memory, and the LAN connection path between two host CPU memories.
This causes a long latency and a slow data throughput.

In the lattice QCD simulations the most compute and communication intensive part is the multiplication of the lattice Dirac operator on fermion fields in the linear equation (quark) solver:
\begin{equation} 
  D \phi = \eta,
\label{eq:EPD}
\end{equation}
where $D$ is a lattice Dirac operator, $\phi$ and $\eta$ are fermion fields. 
To solve Eq.(\ref{eq:EPD}), the Krylov subspace solver algorithms, such as CG, BiCGStab etc., have been used. 
The naive use of the Krylov solver requires many multiplication of $D$ on a vector $v$ such as $w = D v$ to obtain the solution vector $\phi$.
The GPU acceleration can be applied to this operation.
The bottleneck explained above, however, degrades the performance of $w = D v$ operation.
Thus the algorithmic reconsideration is required to remove the bottleneck.

In this paper we study the additive Schwarz domain-decomposition preconditioner with/without domain overlapping~\cite{Smith:2004}.
The additive Schwarz preconditioner is a kind of domain-decomposition preconditioner for elliptic partial differential equations.
L\"uscher has introduced the Schwarz alternating method to the Wilson-Dirac quark solver as the preconditioner and obtained enormous speed up combined with the single precision acceleration technique~\cite{Luscher}.
The Schwarz alternating method corresponds to the multiplicative Schwarz preconditioner and we expect a similar improvement for the additive Schwarz method.
We study this possibility with the $O(a)$-improved Wilson quark on a moderate size lattice.
The performance is measured on a GPU cluster we developed.

\begin{figure}[th]
  \vspace*{-1.0em}
  \centering
  \includegraphics[width=0.9\textwidth]{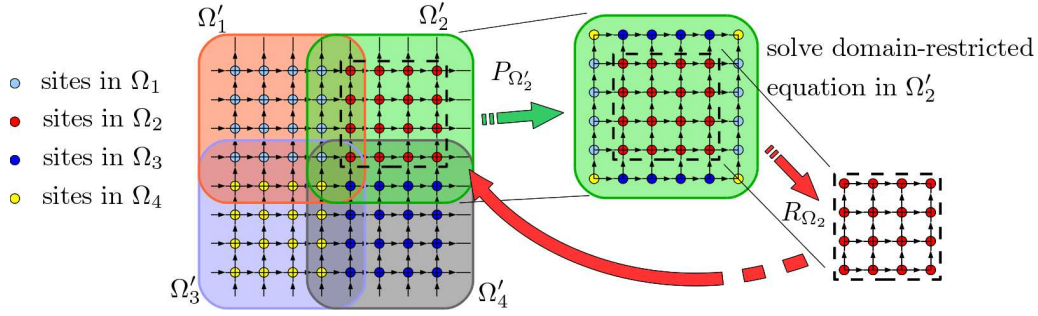}
  \vspace*{-0.5em}
  \caption{Lattice domain-decomposition and relation to the RAS iteration.}
  \label{fig:RASDecompLat}
  \vspace*{-0.5em}
\end{figure}

In the next section, we explain the details of the restricted additive Schwarz (RAS) domain-decomposition iteration~\cite{Cai:1999}. 
To accelerate the solver using multiple GPUs we employ the mixed-precision nested-BiCGStab solver~\cite{Clark:2010,NestedBiCGStab,Buttari:2008}, and we apply the RAS method to the GPU side BiCGStab solver as the preconditioner.
The acceleration of the solver with the GPU and the RAS is explained in section~\ref{sec:BiCG}.
We show the details of our GPU machine and the programming environment in section~\ref{sec:Programing}.
We test the effect of the RAS preconditioner varying the parameters of the RAS preconditioner and study the bottleneck by investigating the timing chart 
of the algorithm.
The results are shown in section~\ref{sec:results} and we give a brief summary for the results in the last section.

\section{The Restricted Additive Schwarz domain-decomposition iteration} 
\label{sec:RAS}

The restricted additive Schwarz iteration~\cite{Cai:1999} 
is a kind of the fixed iteration solver for elliptic differential equations.
This solver makes use of the geometrical structure of a latticized partial difference equation.
In lattice QCD the discretized space-time can be split into several domains and we show 
the schematic picture of the decomposition in Fig.\ref{fig:RASDecompLat}.
$\Omega_i$ represents the lattice sites in the $i$-th domain without overlapping.
$\Omega'_i$ denotes the domain extended from $\Omega_i$.
The extended domains are overlapped in general and the data in overlapped 
region are replicated on the neighbouring domains.

To solve Eq.~(\ref{eq:EPD}) without domain overlapping, we expect that 
the solution $\phi$ can be approximated by combining the partial solution of
$\xi_{\Omega_i}$ derived from $D_{\Omega_i} \xi_{\Omega_i} = \eta_{\Omega_i}$ 
from each domain,
where $D_{\Omega_i}$ is the restriction of $D$ to $\Omega_i$ with 
the Dirichlet boundary condition.
The additive Schwarz (AS) iteration simply approximates it as
$\phi \sim \sum_i \xi_{\Omega_i}$, and the approximation is refined 
by the Richardson iteration.
A problem arises when we overlap the decomposition
since the approximate solution derived from the extended equation
$D_{\Omega'_i} \xi_{\Omega'_i} = \eta_{\Omega'_i}$ becomes inconsistent in the overlapped region.
The restricted additive Schwarz (RAS) iteration
gives a simple solution to this inconsistency.
In Fig.\ref{fig:RASDecompLat} we denote the restriction operation as $R_{\Omega_i}$ arrow 
which simply extracts the data on the bulk sites ($\Omega_i \in \Omega'_i$) 
to avoid the inconsistency.
Thus the approximation to $\phi$ can be constructed as $\phi \sim \sum_i R_{\Omega_i} \xi_{\Omega'_i}$.
We show the RAS iteration in Alg.~\ref{alg:RAS}.
The fourth line pickups the data on $\Omega'_i$ from the whole field vector, 
the fifth line solves the target problem restricted in the overlapped domain $\Omega'_i$ with 
the Dirichlet boundary condition, and the next line represents the restriction process described above.

\begin{figure}[t]
  \vspace*{-2em}
  \begin{algorithm}[H]
    \caption{The RAS iteration. This calculates $\phi = K_{\rm RAS} \eta \sim D^{-1} \eta$ 
     with {\it input} $\eta$, and {\it output} $\phi$.}
    \label{alg:RAS}
    \begin{algorithmic}[1]
      \STATE set initial solution to $\phi = 0$ and $r = \eta$.
      \FOR{$j = 0, 1, \ldots, NRAS - 1$}
        \FOR{$i = 1, \ldots, N$ (domain block index loop)}
           \STATE $r_{\Omega'_i} = P_{\Omega'_i} r$. (Projection to domain $\Omega'_i$.)
           \STATE solve $D_{\Omega'_i} v_{\Omega'_i} = r_{\Omega'_i}$ for $v_{\Omega'_i}$
                  with the Dirichlet boundary condition.
           \STATE $v_{\Omega_i} = R_{\Omega_i} v_{\Omega'_i}$. (Restriction to domain $\Omega_i$.)
        \ENDFOR
        \STATE $v = \sum_i^N v_{\Omega_i}$.
        \STATE update $\phi = \phi + v; r = r - D v$.
      \ENDFOR
    \end{algorithmic}
  \end{algorithm}
  \vspace*{-1em}
\end{figure}

The RAS iteration itself is not sufficient for the complete solver, and 
is usually used as the preconditioner for the Krylov subspace iterative solvers.
We employ BiCGStab solver for the Krylov subspace solver.
The RAS preconditioner $K_{\rm RAS}$ corresponds to the following operator;
\begin{equation}
  K_{\rm RAS} = S \sum_{j=0}^{NRAS-1} (1-DS)^j, \quad\mbox{with}\quad    
            S = \sum_{i=1}^N  R_{\Omega_i} ( D_{\Omega'_i}^{-1} ) P_{\Omega'_i}.
\end{equation}
This is applied to the following preconditioned equation;
\begin{equation}
  D K_{\rm RAS} \chi = \eta, \quad \phi = K_{\rm RAS} \chi,
\label{eq:precdEQ}
\end{equation}
where $\chi$ is to be solved with the Krylov subspace iterative solvers. 
Note that in the additive Schwarz case, the domain equation can be solved independently 
from other domains and the domain index $i$ in Alg.~\ref{alg:RAS} can be completely 
parallelized. We assign a single domain to a single GPU in this paper.

To obtain the best performance with the RAS preconditioner 
we should appropriately optimize the following three parameters. 
The one is the depth of the overlapped region $d$.
In Fig.~\ref{fig:RASDecompLat} we show the depth $d=2$ case.
The exact inversion in the individual domain is not required in the RAS preconditioner 
and a fixed iteration solver
with the iteration number $N_{\mathrm{dominv}}$ is usually used for $(D_{\Omega'_i})^{-1}$.
The last parameter is $NRAS$.
These parameters are surveyed in the benchmarking test.

The RAS preconditioner is a kind of generalized (blocked) Jacobi preconditioner.
One can expect that the performance of the RAS becomes more better as increasing the overlap depth $d$
since the equation in a domain approaches to the original equation.
However it requires extra works on the overlapped region and degrades the total performance.
Therefore there should be a optimal choice for $d$.
The domain size also affects the performance.
The larger domain size is more better as the preconditioner but it reduces the domain parallelism.
Overlapping domains has a gain when it is used for the GPU acceleration
because GPU can keep a high performance for larger domain size.
In the next section we explain the details of the GPU implementation of 
the BiCGStab solver and the RAS preconditioner.

\section{Accelerating Krylov solver with the Schwarz method and GPUs} 
\label{sec:BiCG}

The GPU architecture is originally dedicated for computer graphics application and shows a great performance for the single-precision arithmetic.
We employ the mixed-precision nested-BiCGStab (flexible BiCGStab) algorithm~\cite{NestedBiCGStab,Buttari:2008}
to extract the single-precision performance efficiently.
The mixed-precision nested-BiCGStab consists of an inner and an outer BiCGStab solvers, where
the outer solver solely runs with the double-precision while the inner solver works with 
the single-precision.
The inner solver corresponds to the preconditioner to the outer solver and 
the most of arithmetics are done within the inner solver side. 
Thus we can extract the best performance of GPU by assigning the inner solver task 
to GPUs.

As described in the introduction the data communication is a bottleneck of the GPU 
accelerated parallel computing.
There is a possibility for the RAS preconditioner to reduce the bottleneck by appropriately 
matching the domain-decomposition and the node allocation.
In this paper we split the lattice so that a single GPU is responsible to a single domain (block)
and assign the RAS preconditioner to the inner single-precision BiCGStab solver.
In this manner we can extract the true performance of GPUs since the domain equations of 
the RAS preconditioner are solved in completely parallel without 
data communication.

The data communication arises in the 4th and 9th lines of Alg.~\ref{alg:RAS}.
The $D v$ operation of the 9th line is always required for any iterative solvers 
and the 4th line (projection to $\Omega'_i$) is the extra communication 
arises in the RAS preconditioner.
The performance gain from the RAS preconditioner is expected when
the total iteration count for the inner solver is sufficiently reduced by the RAS preconditioner
so as to beat the increase of the communication overhead from the projection operation.

\section{Machine and programinng}
\label{sec:Programing}

We construct a GPU cluster consists of four PC boxes. 
Each PC has a Intel Core i7 920 running at 2.67GHz, 6GBytes DDR3 memory,
two GeForce GTX 285 GPU cards, and a single Intel Gigabit ET Quad Port Server adapter.
The Gigabit Ethernet connection is rather slow 
but we make use of the four ports via the network trunking facility of the OpenMPI libraly.
The operating system is CentOS 5.3. The programing language we employed is 
Intel Fortran for the outer double-precision BiCGStab solver (CPU host code) and 
NVIDIA CUDA 2.3 for the inner single-precision BiCGStab solver and the RAS preconditioner (GPU code).
To further improve the Gigabit Ethernet performance, we use Open-MX protocol,
which is freely available from~\cite{Open-MX}, instead of TCP/IP.

We split the whole lattice with the size of $N_x N_y N_z N_t$
into $N_{\mathrm{GPU}}=8$ domains by dividing $x$-direction only.
We extend the domain size by adding extra ghost/overlap region in both upward and downward 
$x$-directions to constact the overlapped domain-decomposition.
The resulting domain size becomes $(N_x/N_{\mathrm{GPU}}+2s)N_y N_z N_t$ where
$s$ is the extension size and the depth of the overlap is $d=2s$.
This one-dimensional splitting is preferable compared to the multi-dimensional splitting 
in view of the communication overhead.
The data structure is important to achieve the best performance of the Nvidia's cards.
As described in Refs.~\cite{Clark:2010,CudaGuide}, we have to carefully arrange 
the data ordering to extract the best performance. 
We assign a single CUDA thread to a single site.
For the details for the CUDA threading/blocking in lattice QCD simulations, see
Ref.~\cite{Clark:2010}.

To clarify the bottleneck of the solver with the RAS preconditioner accelerated by multiple GPUs
we investigate the timing chart of the whole algorithm until the solver yields the double-precision 
solution.
In the next section we will show the computing and communication time for the following region:
$T^{\mathrm{comm}}_{\mathrm{proj}}$ for the communication time at the projection (4-th line of Alg.~\ref{alg:RAS}),
$T^{\mathrm{comm}}_{Dv}$ for the communication time in the Wilson-Dirac operator multiplication 
(9-th line of Alg.~\ref{alg:RAS} and Eq.~(\ref{eq:precdEQ})), 
$T^{\mathrm{calc}}_{Dv}$ the computation time in the Wilson-Dirac operator multiplication, and
$T^{\mathrm{calc}}_{\mathrm{dominv}}$ the computation time in the approximate inversion of the domain-restricted Wilson-Dirac operator.

\section{Results}
\label{sec:results}

We measure the solver time using a random gauge configuration on a $32^4$ lattice.
The block size for a single GPU becomes larger than $4\times 32^3$ 
which is enough size to extract the true performance of the GPU.
The parameters for the $O(a)$-improved Wilson-Dirac fermion are chosen to be $\kappa = 0.126$ 
and $c_{\rm SW} = 1.0$ with which the solver converges after enough iteration for the timing measurement.
The solver stopping condition is $|\eta - D \phi|/|\eta| < 10^{-14}$ with a Gaussian noise vector $\eta$.
We have investigated the performance of the RAS and GPU accelerated solver 
with the RAS parameters ranging in $NRAS = 1$--$10$ and $N_{\mathrm{dominv}} = 1$--$20$ 
and find that the combination of $NRAS = 3$ and $N_{\mathrm{dominv}} = 5$ is 
the best parameter for $d = 0,2,4$.
In this section we show the results with the best parameters only.

\begin{table}[htbp]
\centering
  \begin{tabular}{|c|c|c|c|c|c|} \hline
      \multicolumn{2}{|c|}{ }                    &No prec.&RAS($d=0$)&RAS($d=2$)&RAS($d=4$) \\
      \hline \hline
      \multicolumn{2}{|c|}{total time [sec]}      & 53.305 & 27.977 & 35.234 & 37.195 \\
      \hline \hline
      \multicolumn{2}{|c|}{GPU solver time [sec]} & 52.380 & 26.254 & 33.540 & 35.197 \\
      \hline
      \multicolumn{2}{|c|}{copy time [sec]}       & 47.358 & 17.917 & 24.883 & 25.880 \\
      \hline
      \multicolumn{2}{|c|}{$Dv$ operation count}  &  1,328 &    484 &    496 &    416 \\
      \hline
                                    & total [sec] & 48.243 & 18.246 & 18.441 & 15.546 \\
      \cline{2-6}
      $Dv$ time & $T_{Dv}^{\mathrm{calc}}$ [sec]    &  3.900 &  1.424 &  1.436 &  1.205 \\
      \cline{2-6}
                 & $T^{\mathrm{comm}}_{Dv}$ [sec]   & 47.358 & 17.917 & 18.111 & 15.269 \\
      \hline
      \multicolumn{2}{|c|}{$T^{\mathrm{comm}}_{\mathrm{proj}}$ [sec]} &
                                                         - & 0.0    & 6.772  & 10.611 \\
      \hline
      \multicolumn{2}{|c|}{$T^{\mathrm{calc}}_{\mathrm{dominv}}$ [sec]} &
                                                         - & 5.954  & 6.245  & 6.993  \\
      \hline
  \end{tabular}
\caption{Timings for performance comparison.}
\label{tab:results}
\end{table}

Table~\ref{tab:results} shows the results from our benchmarking tests. 
The first column is the result without preconditioning and 
the others are with the RAS preconditioner with $d=0,2,4$ respectively.
The first row shows the timing for the convergence in double-precision and the second row
shows the timing of the GPU solver involved in the total time.
Using the mixed-precision nested BiCGStab algorithm, the most of the computation are done within the GPU solver as expected.
The fastest is obtained with the RAS without domain-overlapping and 
this is against the expectation for the effect of the overlapped
domain-decomposition. 

The copy time represents the timing for the communication which consists of those in the $Dv$ multiplication ($T^{\mathrm{comm}}_{Dv}$) and in the projection ($T^{\mathrm{comm}}_{\mathrm{proj}}$).
The communation time dominates the total time and the bi-directional bandwidth is observed to be $\sim$300 MByte/sec.

The next row counts the $Dv$ operation in the GPU solver. 
With the RAS preconditioner the $Dv$ operation is much reduced from that without preconditioner as expected in section~\ref{sec:BiCG}. 
However overlapping domains does not reduce the $Dv$ operation from $d=0$ to $d=2$, and only a slight reduction is observed in the case from $d=0$ to $d=4$.
The timings involved in the $Dv$ operation are shown in the next row labeled by ``$Dv$ time''.
The $Dv$ operation is dominated by the communication time, although we hide the communication behind the bulk computation of $Dv$.
Therefore the reduction of the $Dv$ operation count is almost identical to the reduction of the communication on our GPU cluster.

The rows labeled by $T_{\mathrm{proj}}^{\mathrm{comm}}$ and $T_{\mathrm{dominv}}^{\mathrm{calc}}$ show the timings for the projection communication and for the approximate inversion of the domain-restricted equation respectively.
These timings are the extra cost for the RAS preconditioner.
From these results we observe that the $Dv$ operation count reduction in the $d=4$ case does not help the total timinig reduction since the extra overhead from the projection and the domain inversion exceeds the gain from the $Dv$ operation reduction. 
The same statement would hold even if there is a slight reduction of the $Dv$ operation count in the $d=2$ case.

\section{Summary}

We have investigated the acceleration of the quark solver using multiple GPUs in parallel with the combination of the RAS preconditioner and the mixed-precision nested-BiCGStab algorithm. 
Parallel GPU benchmarking tests have been done on a GPU cluster constructed for low cost lattice QCD simulations.
The network device is slow compared to the speed of the GPU cards.
Using the RAS preconditioner with the appropriate domain-decomposition and the GPU task assignment, we can reduce the data communication overhead and have observed a factor two improvement with the RAS without domain-overlapping.
However the overlapped domain-decomposition method does not work well on our GPU cluster due to the extra overhead arising from the projection operation and the inversion of the domain-restricted equation. 
The results with the RAS preconditioner is still dominated by the communication time.

A part of the program development has been done on the INSAM (Institute for Numerical Simulations and Applied Mathematics) GPU cluster at Hiroshima University. 
This work was supported in part by the Grant-in-Aid for Scientific Research of Japan Society for the Promotion of Science (JSPS) (No. 20740139).

\end{document}